\documentclass[jgrga]{agutex}
\usepackage{amsmath}
\usepackage{amssymb}
\usepackage[]{graphicx}
\authorrunninghead{Komar and Cassak}
\titlerunninghead{Local Analysis of Dayside Reconnection}
\usepackage{verbatim}
\usepackage{mathrsfs}
\usepackage{color}

\begin{document}
{\it Submitted to J.~Geophys.~Res.~Space Physics February 12, 2016; Revised April 28, 2016} \vskip .125cm

\title{The local dayside reconnection rate for oblique interplanetary
  magnetic fields}
\authors{C.~M.~Komar,\altaffilmark{1,2,3}
  P.~A.~Cassak\altaffilmark{1}}

\altaffiltext{1}{Department of Physics and Astronomy, West Virginia
  University, Morgantown, West Virginia, USA}
\altaffiltext{2}{Geospace Physics Laboratory, NASA Goddard Space
  Flight Center, Greenbelt, Maryland, USA}
\altaffiltext{3}{Department of Physics, The Catholic University of
  America, Washington, District of Columbia, USA}

\authoraddr{C. M. Komar, (colin.m.komar@gmail.com).}

\begin{abstract}

  We present an analysis of local properties of magnetic reconnection
  at the dayside magnetopause for various interplanetary magnetic
  field (IMF) orientations in global magnetospheric simulations.  This
  has heretofore not been practical because it is difficult to locate
  where reconnection occurs for oblique IMF, but new techniques make
  this possible.  The approach is to identify magnetic separators, the
  curves separating four regions of differing magnetic topology, which
  map the reconnection X-line.  The electric field parallel to the
  X-line is the local reconnection rate.  We compare results to a
  simple model of local two-dimensional asymmetric reconnection.  To
  do so, we find the plasma parameters that locally drive reconnection
  in the magnetosheath and magnetosphere in planes perpendicular to
  the X-line at a large number of points along the X-line.  The global
  magnetohydrodynamic simulations are from the three-dimensional
  Block-Adaptive, Tree Solarwind Roe-type Upwind Scheme (BATS-R-US)
  code with a uniform resistivity, although the techniques described
  here are extensible to any global magnetospheric simulation
  model. We find that the predicted local reconnection rates scale
  well with the measured values for all simulations, being nearly
  exact for due southward IMF.  However, the absolute predictions
  differ by an undetermined constant of proportionality, whose
  magnitude increases as the IMF clock angle changes from southward to
  northward.  We also show similar scaling agreement in a simulation
  with oblique southward IMF and a dipole tilt.  The present results
  will be an important component of a full understanding of the local
  and global properties of dayside reconnection.

\end{abstract}

\begin{article}

\section{Introduction}
\label{section::Introduction}
Magnetic reconnection occurs where oppositely directed magnetic field
components undergo a change of topology and subsequently combine
together, resulting in energization of the plasma threaded by the
magnetic field.  The simple model of~\citet{dungey1961, dungey1963}
depicts reconnection occurring at the subsolar magnetopause under due
southward interplanetary magnetic field (IMF) and no dipole tilt; when
the IMF has the opposite orientation, reconnection occurs poleward of
the magnetospheric cusps. Dayside reconnection couples solar wind
plasma to the geomagnetic field and subsequently drives many processes
observed in Earth's magnetosphere: magnetospheric
convection~\citep{dungey1961}, expansion of the polar cap
(see~\citet{boudouridis2005} and references therein), and the
development of plasmaspheric drainage plumes (see~\citet{sandel2003}
and references therein) among many others.

Recently, there has been increased debate regarding what physical
parameters determine the rate at which reconnection proceeds at the
dayside magnetopause, both globally and locally.  As the solar wind is
the primary driver of magnetospheric reconnection, it has been argued
that the dayside reconnection site adjusts to reconnect the magnetic
flux from the solar wind at the global rate it is injected
\citep{axford1984}.  This is quantified by coupling functions of the
solar wind's geoeffectiveness.  \citet{newell2007} reviewed several
models that have been used; all the ones listed only depend on solar
wind plasma parameters, underscoring the long-held belief that
geoeffectiveness is controlled by the solar wind.  More recently, a
theoretical model mapping the solar wind (convective) electric field
to that of the subsolar magnetopause, \textit{i.e.}, the subsolar
reconnection rate, was developed (J.~C.~Dorelli, private
communication, 2015).

Recent arguments, however, suggest that this approach neglects
contributions from the magnetospheric plasma that can affect the
reconnection.  \citet{borovsky2006b} showed that geomagnetic indices
decrease during times when plasmaspheric drainage plumes propagate to
the dayside magnetopause.  It was argued that the plumes mass-load the
reconnection site resulting in slower Alfv\'{e}n speeds, thus slowing
reconnection.  This ``plasmasphere effect'' has been further supported
with total electron content (TEC) observations~\citep{walsh2014a} and
numerical simulations~\citep{borovsky2008a, ouellette2016}.  It has
subsequently been argued that the plasmasphere effect changes the {\it
  local} reconnection rate, and concomitantly some geomagnetic
indices, but does not change the {\it global} reconnection
rate~\citep{lopez2016}.  However, mass-loading the magnetosphere also
decreases the global reconnected flux~\citep{zhang2016}, and it was
argued that plumes can change the global reconnection
rate~\citep{ouellette2016}.  Thus, what sets the local, and even the
global, dayside reconnection rate remains an important and debated
question.

In order to make definitive studies about these questions, one needs
an approach to systematically locate and analyze the local properties
of dayside reconnection.  As we discuss below, there have been such
numerical and observational studies in the past, but they have focused
on cases with essentially due southward IMF and the simulations have
left out the dipole tilt.  This study outlines an approach to this
problem for the more generic and challenging case when the IMF is
neither due southward nor due northward, but rather makes an oblique
angle relative to the geomagnetic field.

One model of {\it local} dayside reconnection that has been tested in
the last decade is based on a scaling calculation of reconnection in
asymmetric systems \citep{cassak2007c}.  The calculation has been
tested in scaling studies of anti-parallel asymmetric reconnection in
two-dimensional (2D) slab geometries in resistive magnetohydrodynamic
(MHD)~\citep{borovsky2007, cassak2007c}, two-fluid Hall
MHD~\citep{cassak2008b, cassak2009c}, and
particle-in-cell~\citep{malakit2010} simulations, and the predictions
perform well.  It is also successful in describing reconnection in 2D
resistive-MHD turbulence \citep{servidio2009}.  However, the original
calculation was only for 2D systems and did not include the effect of
an out-of-plane magnetic field among other assumptions, so it is not
{\it a priori} obvious that such a theory applies to the real 3D
dayside magnetopause.

Attempts have been made to determine its applicability to the 3D
magnetopause in global MHD simulations with due southward IMF and no
dipole tilt~\citep{borovsky2008a, ouellette2014}.  In general, the
reconnection rates measured in the simulations found agreement with
the scaling relations presented in~\citet{cassak2007c}.  When there
was a plasmaspheric plume, the local reconnection rate decreased as
predicted \citep{borovsky2008a}.  We are not aware of any simulation
studies addressing this topic at the dayside magnetopause for oblique
IMF, and this remains beyond the scope of the present study.

Observationally, there have been tests of the theory.  It was found to
be successful in laboratory experiments~\citep{yoo2014, rosenberg2015}
and may explain features of asymmetric X-ray emission from the
footpoints of solar flares \citep{murphy2012}.  In the magnetosphere,
Polar observations of reconnection at Earth's dayside magnetopause
have confirmed the scaling of the theory~\citep{mozer2010}; the events
studied were exclusively for nearly due-southward IMF.  It was shown
observationally that reconnection slows locally when a plume reaches
the dayside reconnection site \citep{walsh2014a, walsh2014b}.
Predictions of the substructure of the diffusion region have been
observed \citep{graham2014, muzamil2014}. The Cluster satellites were
used to compare the theory with multiple events, showing a correlation
between the predictions and the data \citep{wang2015} (though see the
paper for further discussion).

One reason assessing the applicability of asymmetric reconnection
theory at the dayside magnetopause for oblique IMF orientations has
been challenging is that the precise locations on the magnetopause
where reconnection occurs for such situations is not well understood.
Several models to locate reconnection and its orientation exist (see,
{\it e.g.,} \citep{trattner2007, swisdak2007, hesse2013}), but a
recent study found that none of the competing models are perfect for
oblique IMF conditions~\citep{komar2015}.  Knowing exactly where
reconnection occurs is crucial for the questions being addressed here.
An alternate approach is to determine the topology of the magnetic
field and identify precisely the location of dayside reconnection as
regions where different magnetic topologies converge.  The curve that
separates the four topologies (solar wind, terrestrial, and open to
the solar wind on one end and piercing Earth's north or south pole on
the other) is the magnetic separator, which marks the location where
reconnection can occur.  The reconnection X-line, therefore, lies
along the separator.  While formally there are differences between
separators and X-lines because reconnection need not be happening
everywhere along a separator, we tend to see in the magnetospheric
geometry that reconnection does happen along most of the separator, so
for the purposes of this paper we often use the words interchangeably.
There now exist numerous methods to determine magnetic separators in
global magnetospheric observations~\citep{xiao2007, pu2013} and
simulations~\citep{laitinen2006, hu2007, komar2013, stevenson2015a,
  glocer2016}, as well as in the solar context~\citep{longcope1996,
  close2004, beveridge2006}.  This approach to finding reconnection
sites has become practical and therefore is the approach used in this
study.
  
In order to systematically study the applicability of asymmetric
reconnection theory to the dayside magnetopause with obliquely
oriented IMF and dipole tilt, one must carefully analyze the
reconnection physics local to the X-line.  We adopt an approach
similar to that of~\citet{parnell2010} for the solar context.  Having
previously located magnetic separators in our global magnetospheric
resistive-MHD simulations with obliquely oriented
IMF~\citep{komar2013} and non-zero dipole tilt~\citep{komar2015}, we
quantify local properties of reconnection on dayside portions of the
X-lines for multiple simulations.  We go beyond previous work by
systematically measuring local parameters in planes normal to the
magnetic separator and comparing the measured reconnection rate at the
separator to the predictions of local asymmetric reconnection theory.
We find that the 2D model reproduces the measured local electric
fields along the X-line quite well in the scaling sense, especially
for the previously-studied due southward IMF case where the agreement
is nearly perfect.  However, for oblique IMF, there is an absolute
constant of proportionality not captured by the model which becomes
more significant as the IMF clock angle decreases from $180^\circ$.
The results are similar for systems with a dipole tilt.  We also show
that care must be employed to ensure proper resolution of the
diffusion region for studies of this sort.

The layout of this paper is as follows:
Section~\ref{section::Methodology} provides a brief overview of the 2D
asymmetric reconnection scaling relations that are tested in the
present study, describes our simulation setup, summarizes the method
employed to determine the magnetic separators in our global
simulations, and outlines the systematic approach used to measure the
local plasma parameters that are used to calculate the local
asymmetric reconnection rate from the scaling relations.  Our
simulation results are presented in Sec.~\ref{section::Results}.
Finally, a brief summary of our results and their applications are
discussed in Sec.~\ref{section::Summary}.

\section{Methodology}
\label{section::Methodology}

\subsection{Reconnection Model to Compare with Simulations}
\label{section::Asymmetric_Reconnection_Model}

The analytical model we compare to our simulations assumes upstream
conditions with a magnetospheric plasma of density $\rho_\text{MS}$
and reconnecting magnetic field component of strength $B_\text{MS}$
with magnetosheath plasma having density $\rho_\text{SH}$ and
reconnecting magnetic field strength $B_\text{SH}$.  The asymmetric
reconnection rate $E_{asym}$ scales as \citep{cassak2007c}
%\begin{linenomath*}
\begin{equation}
  E_{asym}\sim\frac{B_\text{MS} B_\text{SH}}{B_\text{MS} +
    B_\text{SH}}c_{A,asym}\frac{2\delta}{L},
  \label{eqn::Asymmetric_Reconnection_Rate}
\end{equation}
%\end{linenomath*}
where
%\begin{linenomath*}
\begin{equation}
  c_{A,asym}^2\sim\frac{B_\text{MS} B_\text{SH}}{\mu_0\rho_{out}}
  \label{eqn::Asymmetric_Outflow_Speed}
\end{equation}
%\end{linenomath*}
is the predicted outflow speed, $\mu_0$ is the permeability of free
space, the predicted outflow density $\rho_{out}$ is
%\begin{linenomath*}
\begin{equation}
  \rho_{out}\sim\frac{\rho_\text{MS}B_\text{SH}+
    \rho_\text{SH}B_\text{MS}}{B_\text{MS}+B_\text{SH}},
  \label{eqn::Asymmetric_Outflow_Density}
\end{equation}
%\end{linenomath*}
and $\delta$ and $L$ are the half-width and half-length of the
dissipation region, respectively.  This prediction makes no assumption
about the dissipation mechanism.

For the special case of resistive reconnection, as is the case for the
simulations in the present study, the reconnection rate
$E_{\eta,asym}$ was shown to scale as~\citep{cassak2007c}
%\begin{linenomath*}
\begin{equation}
  E_{\eta,asym}\sim\sqrt{\frac{\eta c_{A,asym}}{\mu_0 L}B_\text{MS}
    B_\text{SH}},
  \label{eqn::Asymmetric_SP_Reconnection_Rate}
\end{equation}
%\end{linenomath*}
where $\eta$ is the resistivity.

\begin{figure*}[t]
\centering
\noindent\includegraphics[width=39pc]{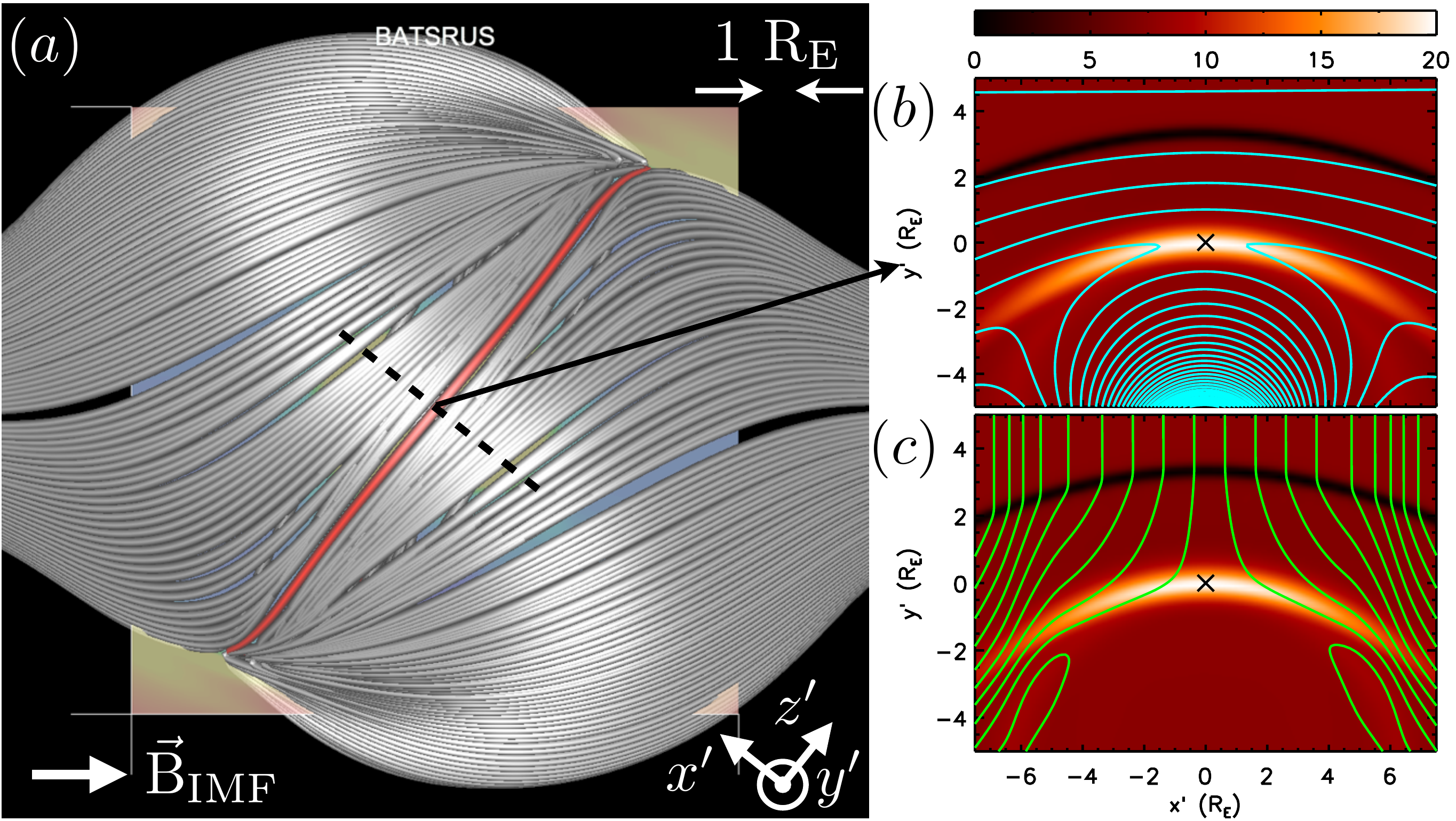}
\caption{The plane normal to the X-line at the subsolar magnetopause
  in a simulation with $\theta_\text{IMF}=90^\circ$.  Panel (a)
  depicts the orientation of the plane centered at
  $\mathbf{r}_{Sep}=\left(8.4,0.0,0.0\right)$~R$_\text{E}$ in GSM
  coordinates (dashed black line).  Panels (b) and (c) display the
  out-of-plane current density $J_{z'}$ in this plane as the color
  background in nA/m$^2$ and displays contours of (b) the flux
  function $\Psi$ in cyan and (c) contours of the stream function
  $\Phi$ in green.  The X-line is located at $\left(0,\,0\right)$ in
  the $x'$-$y'$ plane and marked with an X.}
\label{fig::Example_Separator_Plane}
\end{figure*}

We will test Eqs.~\eqref{eqn::Asymmetric_Reconnection_Rate} and
\eqref{eqn::Asymmetric_SP_Reconnection_Rate} in our global
magnetospheric simulations.  However, it bears noting that there are
limitations to the theory.  It is typically assumed that the magnetic
field component parallel to the X-line, known as the guide field, does
not affect the reconnection rate or dynamics of the dissipation
region.  This is unlikely to be the case in the real system as finite
Larmor radius effects have important consequences for
reconnection~\citep{swisdak2003, swisdak2010, beidler2011,
  malakit2013}.  However, these effects are not present in the
resistive-MHD model used for our global simulation study, so this
assumption may be acceptable for the present study.  Also, the theory
does not include the effect of the solar wind flow on the
magnetosheath side of the magnetopause, which may also be important
\citep{doss2015}.

\subsection{Global Magnetospheric Simulations}
\label{subsection::Global_Magnetospheric_Simulations}

We perform global simulations using the Space Weather Modeling
Framework (SWMF)~\citep{toth2005, toth2012a}, a suite of physical
models developed at the University of Michigan used to model regions
from the sun to the magnetosphere and beyond, although the methods
detailed here can be adapted to other global magnetospheric codes.  We
specifically employ the Block Adaptive Tree Solarwind Roe-type Upwind
Scheme (BATS-R-US) MHD code to solve the resistive-MHD equations on a
high resolution, three-dimensional, rectangular, irregular grid in
order to simulate the global magnetosphere~\citep{powell1999,
  gombosi2000, dezeeuw2000}. The ionosphere is modeled with the
ionospheric electrodynamics (IE) component.
  
The simulations are performed at NASA's Community Coordinated Modeling
Center (CCMC), a freely available code repository administered by NASA
Goddard Space Flight Center.  The CCMC's Kameleon software suite,
which was developed by the CCMC to analyze the standardized output of
different simulation models performed at the CCMC, is used to
partially analyze simulation output of BATS-R-US by sampling data and
tracing magnetic field lines at arbitrary coordinates within the
simulation domain.  The simulation domain is $-255<x<33$, $-48<y<48$,
and $-48<z<48$, where distances are measured in Earth radii
(R$_\text{E}$) and the coordinate system is Geocentric Solar Magnetic
(GSM).  The simulations are run using BATS-R-US version 8.01.

The simulations are evolved for two hours (02:00:00) of magnetospheric
time and we look at the 02:00:00 mark of simulation data when the
dayside magnetopause has achieved a quasi-steady state. This is
determined by comparing the location of the current density $J_y$
along the $x$-axis at adjacent time outputs (every 00:10:00); we find
the current layer along the $x$-axis is approximately stationary after
01:30:00 of magnetospheric time. The standard high-resolution grid for
CCMC simulations has $1,958,688$ grid cells with a coarse resolution
of 8~R$_\text{E}$ in the far magnetotail, and a fine resolution of
0.25~R$_\text{E}$ near the magnetopause. The simulations presented
here employ a maximum resolution of 0.125~R$_\text{E}$ throughout the
region $-15<x,\,y,\,z<15$~R$_\text{E}$, totaling $16,286,400$ grid
cells.

The boundary condition at $x=33$~R$_\text{E}$ is constant solar wind
values, although BATS-R-US is capable of using event data measured by
solar wind monitors.  We use solar wind temperature
$T_\text{SW}=232,100$ K (20~eV), IMF strength $B_\text{IMF} =$ 20~nT,
number density $n_\text{SW}=20$~cm$^{-3}$, and a solar wind velocity
of $\mathbf{v}_\text{SW}=-400$~km/s~$\mathbf{\hat{x}}$ unless
otherwise noted.  These values for $n_\text{SW}$ and $B_\text{IMF}$
are somewhat high, but this enables the high resolution region at the
dayside magnetopause to not be as large.  We also have investigated
simulations with lower $B_\text{IMF}$.  We perform distinct
simulations with IMF clock angles $\theta_\text{IMF}=30^\circ$,
60$^\circ$, 90$^\circ$, 120$^\circ$, 150$^\circ$, and 180$^\circ$
(southward). The IMF does not have a $B_x$ component.  For the present
simulations, we additionally employ a uniform explicit resistivity
$\eta/\mu_0=6.0\times10^{10}$~m$^2$/s.  Although the magnetosphere is
known to be collisionless, including an explicit resistivity allows
for reproducible results that are independent of the numerics
\citep{komar2013}.

The IE component of the SWMF uses the currents of the MHD simulation
at 3.5 R$_\text{E}$ to determine the ionospheric currents at a radial
distance of 1.017 R$_\text{E}$ using conservation of electric charge.
Constant Pederson and Hall conductances of 5 mhos are used to
determine the ionospheric electric field $\mathbf{E}$ from these
ionospheric currents at geomagnetic latitude and longitude discretized
into 1$^\circ$ increments resulting in a 181 x 181 spherical grid.

\subsection{Identification of X-lines in Global Simulations}
\label{subsection::Identification_of_Separators}

We employ the separator mapping algorithm of~\citet{komar2013} which
has been shown to reliably trace the dayside portion of X-lines in
global magnetosphere simulations for any IMF orientation and dipole
tilt~\citep{komar2015}.  In the separator tracing algorithm, a
hemisphere is initially centered around a magnetic null.  The
hemisphere's surface, of radius 1~$\text{R}_\text{E}$ for our
purposes, is discretized into a grid.  The magnetic field lines
piercing the hemisphere at each grid point are traced to determine
their magnetic topology.  The approximate location of the separator is
determined by finding where four magnetic topologies meet on the
hemisphere's surface.  Then, another hemisphere is centered at the
determined separator location, and the procedure is iterated to trace
the separator.  The dayside separator is traced from northern to
southern null in this fashion.  An example is shown in
Fig.~\ref{fig::Example_Separator_Plane}(a) for the
$\theta_\text{IMF}=90^\circ$ simulation.  Open field lines are gray
and the X-line is red.

\subsection{Determination of Planes Normal to the X-line}
\label{subsection::Determinination_of_Planes}

The separator tracing algorithm presented in~\citet{komar2013} results
in a number ($\sim$ 30) of locations lying along each X-line.  It is
typically assumed that the plane of reconnection is normal to the
X-line.  However, \citet{parnell2010} analyzed reconnection local to
separators in resistive-MHD simulations; they argued that the plane
containing reconnection can often be the plane oriented normal to the
separator, but is not necessarily a general feature.  For the purposes
of this study, we adopt the assumption that the plane of reconnection
is normal to the X-line.  This assumption fails as one approaches the
nulls, so we caution the reader that this assumption needs further
scrutiny.

We develop a procedure to construct planes normal to the X-line by
defining an orthonormal basis at every point along the X-line.  As a
motivation for the procedure, consider the X-line in
Fig.~\ref{fig::Example_Separator_Plane}(a).  The plane normal to the
X-line at the subsolar point
$\mathbf{r}_{Sep}=\left(8.4,0.0,0.0\right)$ is sketched as the dashed
line.  We define a coordinate system $(x',y',z')$ for this plane.  The
out-of-plane unit vector $\mathbf{\hat{z}}'$ points along the X-line,
{\it i.e.}, along the magnetic field with
$\mathbf{\hat{z}}'=0.62~\mathbf{\hat{y}}+0.78~\mathbf{\hat{z}}$, with
unprimed vectors given in GSM.  We define the $y'$ direction as the
inflow direction and $x'$ as the outflow direction.  For the case
study, the inflow direction at the subsolar point is radially out, so
$\mathbf{\hat{y}}' = \mathbf{\hat{x}}$.  Finally, the in-plane unit
vector completing the orthonormal triplet is defined by
$\mathbf{\hat{x}}'=\mathbf{\hat{y}}'\times\mathbf{\hat{z}}'$.

With this simple case in mind, we now describe the method by which we
determine this coordinate system at arbitrary points on X-lines.  For
the $k$-th location along the X-line, the out-of-plane unit vector
$\mathbf{\hat{z}}'_k$ is tangent to the X-line.  Using a second order
finite difference, this gives
%\begin{linenomath*}
\begin{equation}
  \mathbf{\hat{z}}'_k=\frac{\mathbf{r}_{k-1}-\mathbf{r}_{k+1}}
         {\left|\mathbf{r}_{k-1}-\mathbf{r}_{k+1}\right|},
\label{eqn::Out-of-plane_Component}
\end{equation}
%\end{linenomath*}
where $\mathbf{r}_{k-1}$ and $\mathbf{r}_{k+1}$ are the previous and
subsequent X-line locations, respectively.  We note that one could
think of defining $\mathbf{\hat{z}}'_k$ by the magnetic field
direction $\mathbf{\hat{b}}$ at the X-line, but this definition fails
when the magnetic field parallel to the separator is small, such as
for due southward IMF $\left(\theta_\text{IMF}=180^\circ\right)$.  The
formulation of Eq.~\eqref{eqn::Out-of-plane_Component} guarantees a
meaningful $\mathbf{\hat{z}}'_k$ for {\it any} IMF orientation and
magnetospheric dipole tilt.

The unit vector $\mathbf{\hat{y}}'_k$ in the direction of the inflow
is given by the normal to the magnetopause at $\mathbf{r}_k$.  This is
calculated by finding the projection of $\mathbf{r}_k$ normal to
$\mathbf{\hat{z}}'_k$. Mathematically, this is represented as
%\begin{linenomath*}
\begin{equation}
  \mathbf{\hat{y}}'_k\propto\mathbf{r}_k -
  \left(\mathbf{r}_k\cdot\mathbf{\hat{z}}'_k\right)\mathbf{\hat{z}}'_k.
\label{eqn::Normal_Component}
\end{equation}
%\end{linenomath*}  
Finally, $\mathbf{\hat{x}}'_k$ completes the right-handed triplet by
taking the cross product
%\begin{linenomath*}
\begin{equation}
  \mathbf{\hat{x}}'_k=\mathbf{\hat{y}}'_k\times\mathbf{\hat{z}}'_k.
\label{eqn::Reconnection_Component}
\end{equation}
%\end{linenomath*}

We note that the coordinate system resulting from this process is
similar to the boundary normal (LMN) coordinate system. The three
directions are analogous to their counterparts where
$\mathbf{\hat{x}}'\equiv\mathbf{\hat{L}}$, pointing in the direction
of the reconnecting component of the magnetic field and corresponding
to the reconnection outflow direction,
$\mathbf{\hat{y}}'\equiv\mathbf{\hat{N}}$ is the inflow direction, and
$\mathbf{\hat{z}}'\equiv-\mathbf{\hat{M}}$ is the out-of-plane (guide
field) direction.  We do not employ minimum variance analysis
\citep{sonnerup1967}, however, because it does not always give
appropriate results, especially when there is a guide field.

With this orthonormal basis, the $x'$-$y'$ plane is assumed to be the
plane of reconnection.  Coordinates of locations in this plane are
translated back to GSM coordinates, and Kameleon is used to sample the
plasma number density $n$, thermal pressure $p$, magnetic field
$\mathbf{B}$, plasma flow $\mathbf{u}$, and current density
$\mathbf{J}$ in this plane. Each plane spans $-7.5\le x' \le7.5$ and
$-5\le y' \le5$~R$_\text{E}$ and each direction is sampled in $\Delta
x'=\Delta y'=0.0625$~R$_\text{E}$ increments; the X-line is centered
at $\left(0,\,0\right)$ in each $x'$-$y'$ plane. Finally, the magnetic
field $\mathbf{B}$, plasma flow $\mathbf{u}$, and current density
$\mathbf{J}$ vectors are transformed from GSM coordinates to the
primed planar coordinates at the X-line, {\it e.g.},
$B_{x'}=\mathbf{B}\cdot\mathbf{\hat{x}}'$.

We show the results of this procedure for the simulation with
$\theta_\text{IMF}=90^\circ$ in
Fig.~\ref{fig::Example_Separator_Plane}. Panels (b) and (c) display
the out-of-plane current density component $J_{z'}$ as the color
background in nA/m$^2$.  The X-line's location in the $x'$-$y'$ plane
is marked with an X at $\left(0,0\right)$.

In order to gain insight into what reconnection might look like in
this plane, we employ a method used in 2D geometries based on the flux
function to determine the structure of the magnetic field.  It is not
formally generalizable to 3D, but this is carried out only for
perspective and no conclusions are drawn from the results. If we
ignore any dependence in the $z'$ direction, then we can define a flux
function $\Psi\left(x',y'\right)$ as
%\begin{linenomath*}
\begin{equation}
  \mathbf{B}=-\mathbf{\hat{z}}'\times\nabla'\Psi,
\label{eqn::Flux_Function}
\end{equation}
%\end{linenomath*} 
where the magnetic field $\mathbf{B}$ and derivatives $\nabla'$ are
only considered in the $x'$-$y'$ plane.  Lines of constant $\Psi$
represent the projection of magnetic field lines into the plane.  The
projected magnetic field lines are depicted in cyan in
Fig.~\ref{fig::Example_Separator_Plane}(b).  We similarly define a 2D
stream function $\Phi$ to obtain the streamlines (field lines of the
velocity vector) in the $x'$-$y'$ plane with the simple substitution
of $\Phi$ for $\Psi$ and bulk velocity ${\bf u}$ for ${\bf B}$ in
Eq.~\eqref{eqn::Flux_Function}.
Figure~\ref{fig::Example_Separator_Plane}(c) displays contours of
constant $\Phi$ in green which give the in-plane streamlines.

Figure~\ref{fig::Example_Separator_Plane} displays several features
that are qualitatively consistent with the local picture of 2D
Sweet-Parker collisional reconnection~\citep{parker1957, sweet1958},
and is consistent with the field structures described
in~\citet{parnell2010}, albeit occurring at the dayside magnetopause
with a dipolar magnetic field. First, the out-of-plane current layer
is elongated.  The reconnecting magnetic field components are
oppositely directed with the IMF pointing along $-\mathbf{\hat{x}}'$
and is carried along $-\mathbf{\hat{y}}'$ in the magnetosheath; the
terrestrial magnetic field points along $+\mathbf{\hat{x}}'$ and
slowly convects towards the magnetopause along $+\mathbf{\hat{y}}'$.
These magnetic fields undergo reconnection at the X-line with newly
reconnected magnetic flux located downstream of the X-line, and
displaying a curved X-point reconnection geometry.  Lastly, the plasma
convects horizontally outward from the X-line along $y'$ with speeds
$\left|\mathbf{u}\right|\approx205$~km/s, higher than the vertically
directed magnetosheath flow speed
$\left|\mathbf{u}\right|\approx150$~km/s.  This suggests reconnection
has a role in accelerating the outflowing plasma.  Thus the plane
normal to the X-line {\it at the subsolar magnetopause} qualitatively
resembles 2D pictures of reconnection.

We note that using Eq.~\eqref{eqn::Flux_Function} to determine
magnetic field lines and streamlines in planes normal to the
reconnection line is only rigorously valid for 2D systems, so it
should not be expected to be valid for arbitrary conditions.  It
likely works remarkably well for the plane we show because of the high
degree of symmetry at the subsolar magnetopause in this simulation.
We point out, however, that none of the subsequent analysis is
dependent on the fields determined in this way; it is merely being
shown here to illustrate that the magnetic field and flow in planes
normal to the reconnection line reasonably appear like those of 2D
reconnection models.

\subsection{Measuring Plasma Parameters in Planes Normal to the X-line}
\label{subsection::Local_Reconnection_Analysis}

To analyze the reconnection in each plane and compare to the theory in
Sec.~\ref{section::Asymmetric_Reconnection_Model}, we need the plasma
parameters just upstream of the dissipation region.  We start by
sampling the out-of-plane current density $J_{z'}$ along
$\mathbf{\hat{y}}'$ at $x'=0$ to determine the location of maximum
current density $J_{max}$. Note that $J_{max}$ may not be located at
the X-line and can be offset during asymmetric reconnection at the
dayside magnetopause~\citep{cassak2007c, komar2015}. We define the
full-width, half-max of $J_{z'}$ as the dissipation region's thickness
2$\delta$. We define $y'_\text{SH}$ and $y'_\text{MS}$ as the
locations corresponding to the magnetosheath and magnetospheric edges
where the current is $0.5J_{max}$.  The magnetosheath and
magnetosphere pressures, densities, and the flow and magnetic field
components are measured at $\left(0,\,y'_\text{SH}+\delta\right)$ and
$\left(0,\,y'_\text{MS}-\delta\right)$, respectively, in the $x'$-$y'$
plane.

\begin{figure}[t]
\centering 
\noindent\includegraphics[width=20pc]{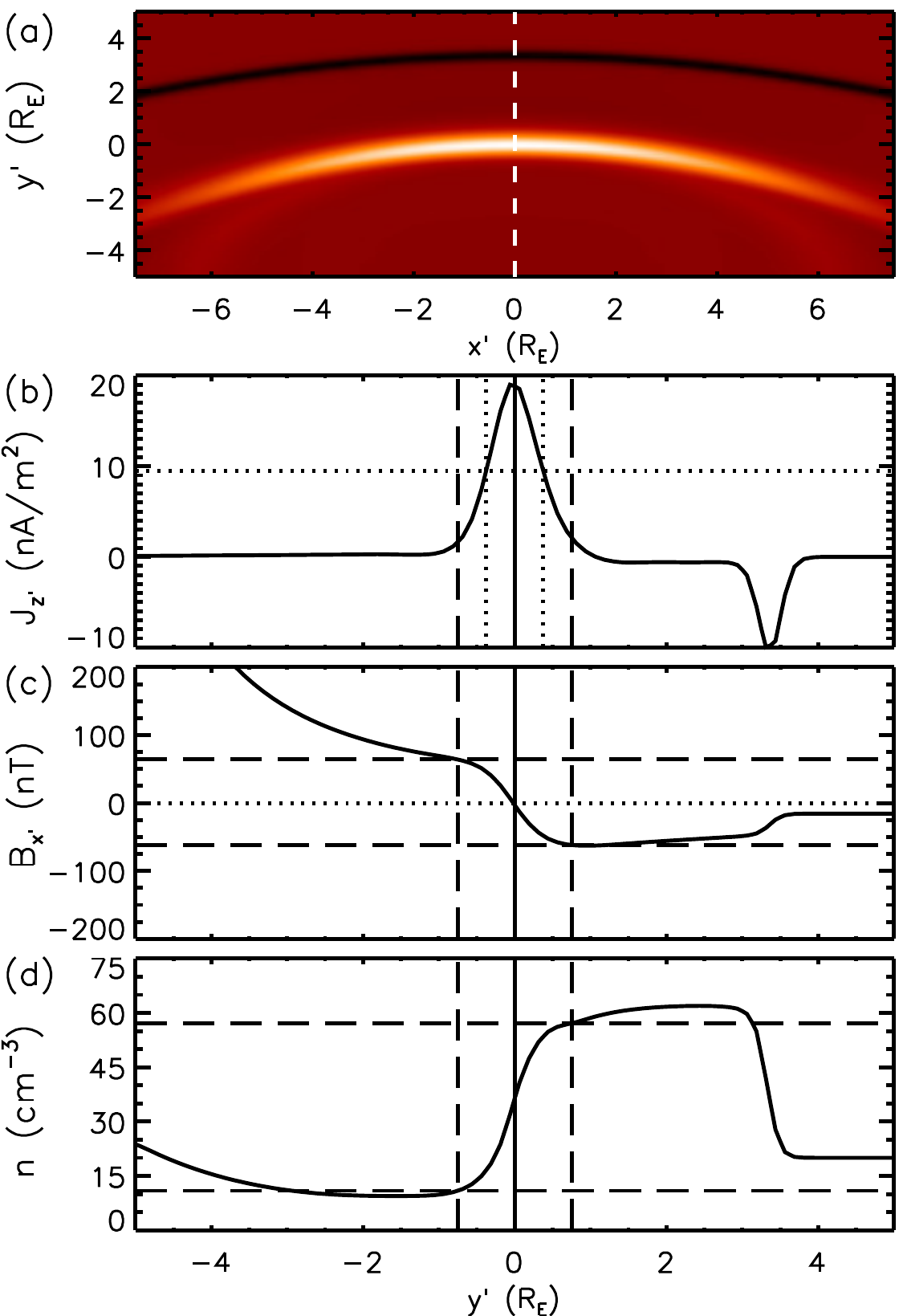}
\caption{Determination of the upstream plasma parameters in the
  simulation with $\theta_\text{IMF}=90^\circ$ for the plane at the
  subsolar point from Fig.~\ref{fig::Example_Separator_Plane}.  (a)
  Out-of-plane current density $J_{z'}$; the white dashed line at
  $x'=0$ displays the line along which plasma parameters are sampled.
  (b) $J_{z'}$ along $x'=0$ with the X-line's location depicted by the
  solid line.  Locations where $J_{z'}=0.5J_{max}$ are marked by
  vertical dotted lines. The value of $0.5J_{max}$ is marked by the
  horizontal dotted line. The vertical dashed lines in panels (b)-(d)
  indicate where magnetospheric and magnetosheath parameters are
  measured.  (c) Reconnecting magnetic field component $B_{x'}$ in nT.
  (d) Plasma number density $n$ in cm$^{-3}$.  The horizontal lines in
  (c) and (d) mark the magnetospheric and magnetosheath values of
  these parameters.}
\label{fig::Measuring_Upstream_Parameters}
\end{figure}
An example of this procedure is demonstrated in
Fig.~\ref{fig::Measuring_Upstream_Parameters}, which is the same plane
displayed in Fig.~\ref{fig::Example_Separator_Plane}.  Panel (a)
displays the out-of-plane current density $J_{z'}$ as the color
background with a dashed white line at $x'=0$, the line along which
the plasma parameters are sampled. Panel (b) shows the out-of-plane
current density along $x'=0$, with vertical dotted lines at the
locations $y'_\text{MS}$ and $y'_\text{SH}$ where the current density
has the value $0.5J_{max}$, marked with a horizontal dotted line. The
X-line's location is marked with a solid vertical line at $y'=0$.  The
left and right dashed vertical lines mark $y'_\text{MS}-\delta$ and
$y'_\text{SH}+\delta$ where the magnetospheric and magnetosheath
plasma parameters are measured, respectively.  Panel (c) displays the
reconnecting magnetic field component $B_{x'}$ in nT and panel (d)
displays the plasma number density $n$ in cm$^{-3}$, respectively,
along the same cut.  The locations where the upstream parameters are
sampled are again displayed with vertical dashed lines.  Dashed
horizontal lines in panels (c) and (d) display the values obtained
from this analysis. One can see that each determined parameter is
representative of the asymptotic regions upstream of the dissipation
region, as desired.  The upstream values for this plane on the
magnetospheric side are $B_{x',\text{MS}}=64$~nT and
$n_\text{MS}=11$~cm$^{-3}$ and for the magnetosheath plasma are
$B_{x',\text{SH}}=-61$~nT, and $n_\text{SH}=57$~cm$^{-3}$; the
dissipation region has half-width $\delta=0.38$~R$_\text{E}$.

\begin{figure}[t]
\centering
\noindent\includegraphics[width=20pc]{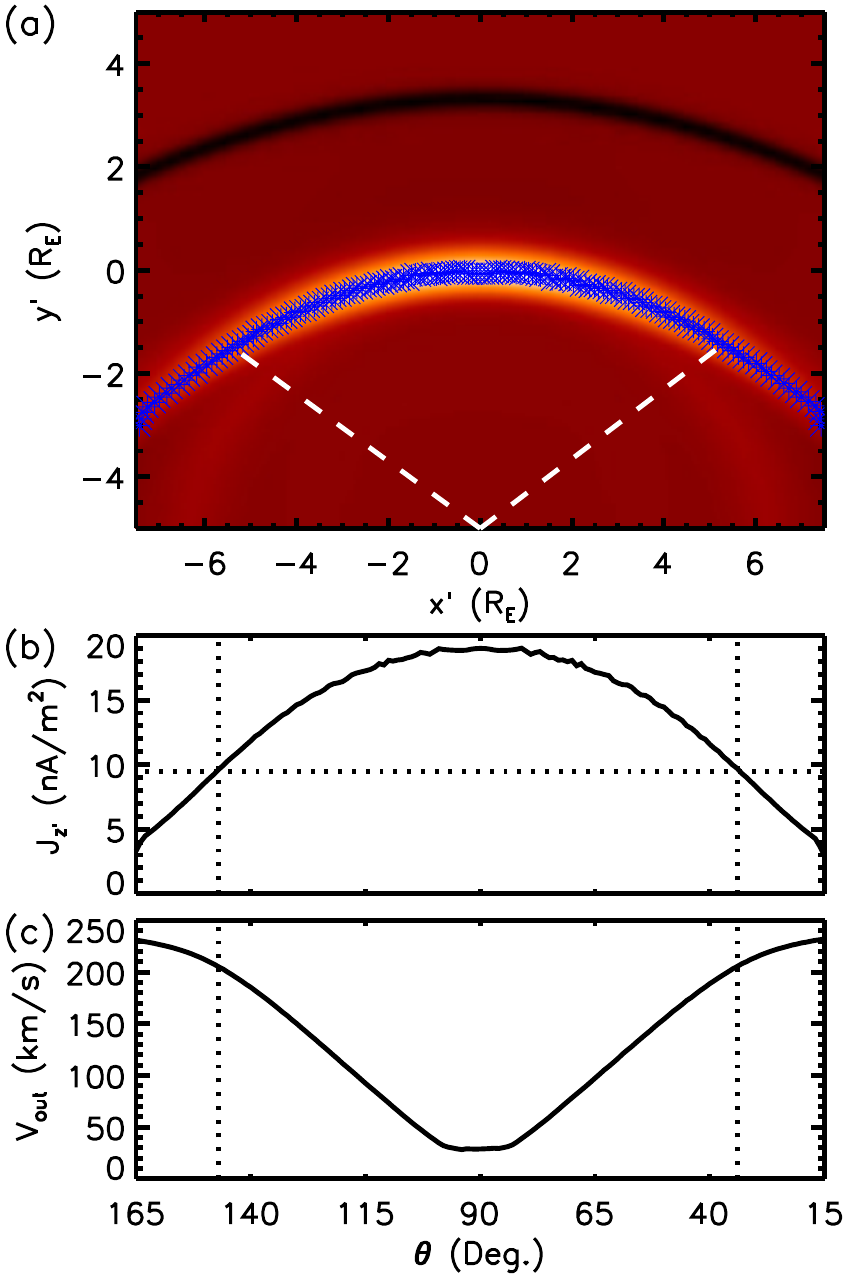}
\caption{Determination of the dissipation region half-length $L$ and
  outflow speed $V_{out}$ for the plane in
  Fig.~\ref{fig::Example_Separator_Plane}.  (a) Out-of-plane current
  density $J_{z'}$.  The locations of current maxima are displayed as
  blue asterisks.  (b) Current density maxima in nA/m$^2$ and (c)
  magnitude of $V_{out}$ in km/s as functions of sampling angle
  $\theta$ measured from the $+x'$ axis.  The left and right edges
  $\theta_{Left}$ and $\theta_{Right}$ of the dissipation region are
  displayed as dotted vertical lines in (b) and (c) and are determined
  by where the current density maximum falls to $0.5J_{max}$,
  indicated by the horizontal dotted line in (b).}
\label{fig::Sample_Dissipation_Region}
\end{figure}

In order to check the validity of the asymmetric reconnection scaling
relations, we must also determine the half-length $L$ of the
dissipation region [see Eqs.~(\ref{eqn::Asymmetric_Reconnection_Rate})
  and \eqref{eqn::Asymmetric_SP_Reconnection_Rate}].  Care must be
taken in determining the dissipation region length as it is curved due
to the structure of Earth's magnetosphere.  We therefore start by
identifying the reconnection dissipation region by sampling the
out-of-plane current density $J_{z'}$ along cuts oriented at an angle
$\theta$ from the $+x'$ axis in the $x'$-$y'$ plane as displayed in
Fig.~\ref{fig::Sample_Dissipation_Region}(a); the cuts start at
$\left(0,-5\right)$~R$_\text{E}$ and the current density is sampled in
1/16 R$_\text{E}$ increments, with $\theta$ discretized into 1$^\circ$
increments from $\left[0^\circ,\,180^\circ\right]$.  The location and
value of the first current density maximum along each cut is retained.
The right and left edges of the dissipation region are defined as
$\theta_{Right}$ and $\theta_{Left}$ where the current density maximum
first achieves a value of $J_{z'}=0.5J_{max}$, where $J_{max}$ is the
aforementioned maximum current density value along $x'=0$. $L$ is
directly calculated from the arc length of the measured current
density maxima locations as
%\begin{linenomath*}
\begin{equation}
  2L=\int_{\theta_{Right}}^{\theta_{Left}}dS\simeq
  \sum_{j=\theta_{Right}}^{\theta_{Left}}\Delta S_j,
\label{eqn::Dissipation_Half_Length}
\end{equation}
%\end{linenomath*}
where $\Delta S_j$ is the distance between the $j$th current density
maximum at $\mathbf{S}_j$ and its neighbor at $\mathbf{S}_{j+1}$ given
by
%\begin{linenomath*}
\begin{equation*}
  \Delta S_j = \left|\mathbf{S}_{j+1}-\mathbf{S}_{j}\right|.
\end{equation*}
%\end{linenomath*}

The outflow speed $V_{out}$ is sampled separately along cuts in
$\theta$.  The outflow speed maxima occur consistently on the
magnetospheric side of the dissipation region, consistent with
previous 2D asymmetric reconnection simulations which measured the
outflows on the side with higher Alfv\'en speed~\citep{cassak2007c,
  birn2008}. The left and right measured outflow velocities are both
205~km/s. We note that the outflow speeds in each direction are
identical at the subsolar point, but they are not in planes away from
the subsolar point.  This asymmetric outflow could be related to
differences in the outflow pressures which has been shown to affect
the outflow speeds~\citep{murphy2010}.

Figure~\ref{fig::Sample_Dissipation_Region} displays the results of
this current density sampling method for the plane normal to the
X-line at the subsolar magnetopause in the
$\theta_\text{IMF}=90^\circ$ simulation.  Panel (a) displays the
current density maxima as blue asterisks. Panel (b) displays current
density maxima values in nA/m$^2$ and (c) displays the outflow
velocity $V_{out}$ in km/s as functions of the sampling angle
$\theta$.  Vertical dotted lines display the determined locations of
$\theta_{Left}$ and $\theta_{Right}$, with the horizontal dotted line
displaying $0.5J_{max}$.  The dissipation region's half-length
$L=5.84$~R$_\text{E}$ for this plane.

\begin{figure*}[t]
  \centering
  \noindent\includegraphics[width=39pc]{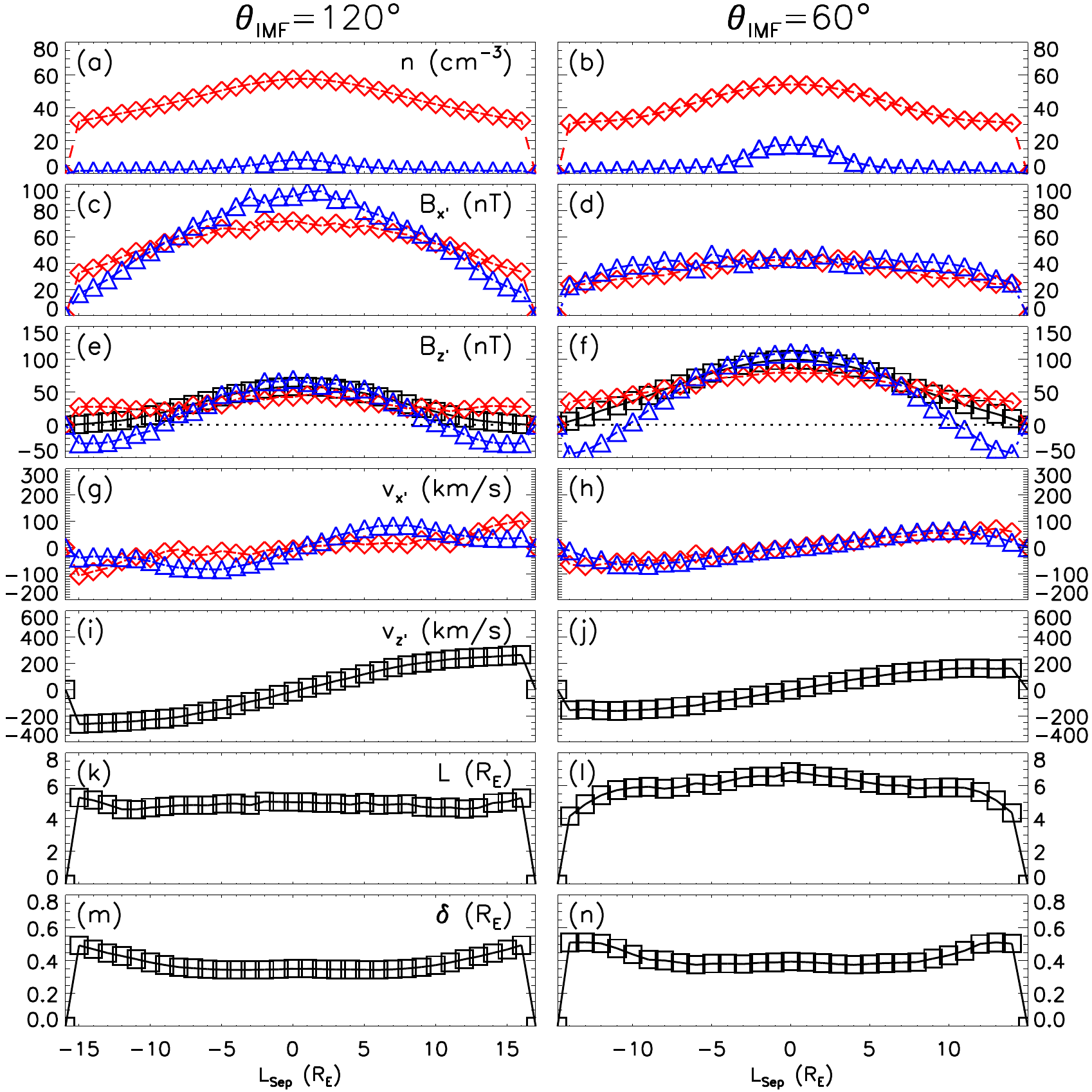}
  \caption{Plasma parameters at the magnetopause (black $\Box$), on
    the magnetospheric side (blue $\triangle$) and on the
    magnetosheath side (red $\Diamond$) obtained with the techniques
    described in the text.  The left and right columns are for
    $\theta_\text{IMF}=120^\circ$ and 60$^\circ$, respectively.
    Displayed in each panel are: (a)-(b) plasma number density $n$ in
    cm$^{-3}$; (c)-(d) reconnecting magnetic field component $B_{x'}$
    in nT; (e)-(f) out-of-plane (guide) magnetic field component
    $B_{z'}$ in nT; (g)-(h) flow parallel to the reconnecting magnetic
    field component $u_{x'}$ in km/s; (i)-(j) out-of-plane flow
    $u_{z'}$ in km/s; (k)-(l) half-length $L$ of the dissipation
    region in $R_E$; and (m)-(n) half-width $\delta$ of the
    dissipation region.  All parameters are displayed as functions of
    $L_{Sep}$, the duskward distance along the X-line from the
    subsolar point.}
  \label{fig::60_120_Parameters}
\end{figure*}
We have now measured all of the relevant parameters to make a
meaningful comparison with the theoretical asymmetric reconnection
scaling relations and the reconnection rate measured in our global
simulations.  We give two examples of the upstream parameters obtained
from this approach in Fig.~\ref{fig::60_120_Parameters}, which show
the sampled plasma parameters in our simulations with
$\theta_\text{IMF}=120^\circ$ (left column) and 60$^\circ$ (right
column) as a function of $L_{sep}$, the duskward distance along the
separator relative to the subsolar magnetopause in R$_\text{E}$;
positive values lie along the northern and dusk flank.  From the
upstream reconnecting magnetic field components $B_{x'}$ and densities
$n$, the dissipation region's half-width $\delta$ and half-length $L$,
we can calculate $E_{asym}$ and $E_{\eta,asym}$ from the asymmetric
scaling relations given by
Eqs.~\eqref{eqn::Asymmetric_Reconnection_Rate} and
\eqref{eqn::Asymmetric_SP_Reconnection_Rate}, respectively.  Both of
these are compared with the local reconnection rate at the X-line
$E_{z'}=\eta J_{z'}$.  The procedure outlined here is repeated for all
planes normal to the X-line for all simulations in this study.

\section{Results}
\label{section::Results}

\begin{figure}[t]
\centering
\noindent\includegraphics[width=20pc]{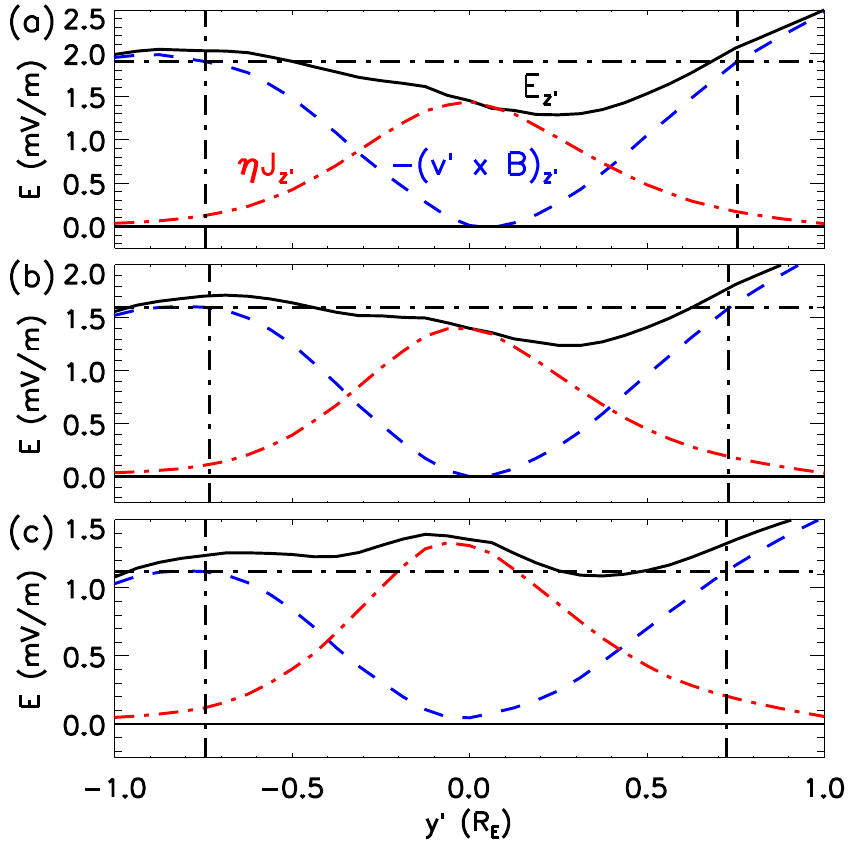}
\caption{Contributions to Ohm's law in three representative planes for
  the simulation shown in Fig.~\ref{fig::Example_Separator_Plane}.
  The panels show the convective (blue dashed), resistive (red
  dot-dashed), and total (black solid) out-of-plane electric field
  $E_{z'}$ in planes at $L_{sep}$ of (a) 0 (b) 5, and (c) 7
  R$_\text{E}$ duskward from the subsolar point.}
\label{fig::ohmslawcuts}
\end{figure}

First, we show the contributions to Ohm's law in representative planes
normal to the X-line to motivate that the results presented here are
reasonable.  Figure~\ref{fig::ohmslawcuts} shows the convective (blue
dashed), resistive (red dot-dashed), and total (black solid) electric
fields in the out-of-plane ($z'$) direction.  The panels are at
$L_{sep}$ of (a) 0 (the subsolar point), (b) 5, and (c) 7 R$_\text{E}$
for the simulations shown in Fig.~\ref{fig::Example_Separator_Plane}.
In calculating the convective electric field, we boost into the
reference frame making the upstream values equal, {\it i.e.,} the
reference frame of the X-line, using a technique used previously
\citep{mozer2002, cassak2009c}.  The vertical dot-dashed lines are the
upstream positions, and the horizontal dot-dashed line is the electric
field at those positions.  The results reveal that the resistive
electric field is nearly equal to the convective electric field in all
three planes, which (1) confirms that the explicit resistivity is
providing the dissipation (rather than numerical effects), (2)
suggests that the steady state assumption is valid, and (3) suggests
the approach we are using to find the plane of reconnection is
reasonable.

\begin{figure*}[t]
\centering
\noindent\includegraphics[width=39pc]{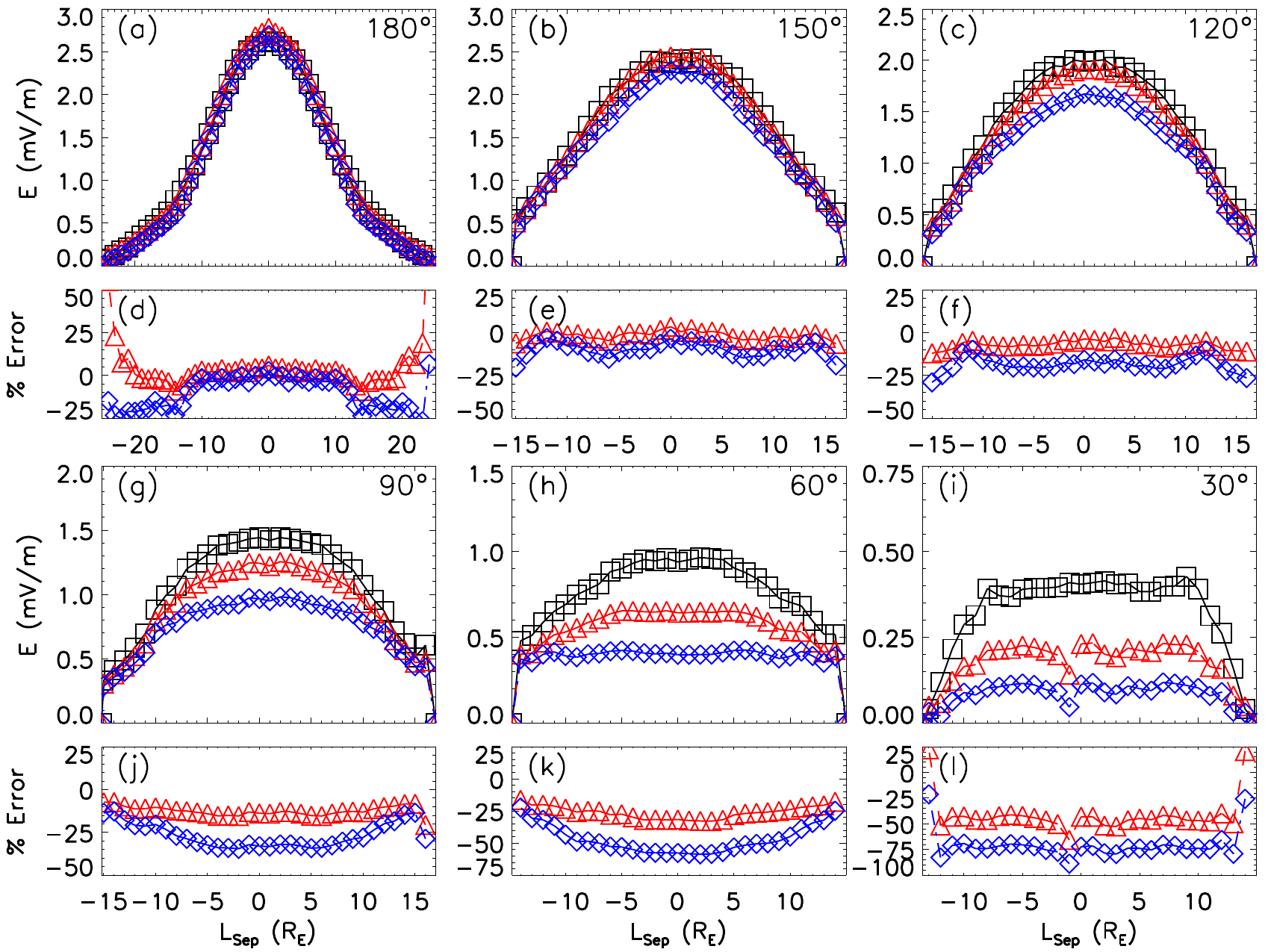}
\caption{Comparison of the measured reconnection rate $E_{z'}$ (black
  $\Box$) at the X-line with the theoretical $E_{asym}$ (blue
  $\Diamond$) and the Sweet-Parker $E_{\eta,asym}$ (red $\triangle$)
  reconnection rates in distinct simulations with $\theta_\text{IMF}$
  of (a) 180$^\circ$, (b) 150$^\circ$, (c) 120$^\circ$, (g)
  90$^\circ$, (h) 60$^\circ$, and (i) 30$^\circ$.  Percent errors
  between $E_{z'}$ and $E_{asym}$ (blue $\Diamond$) or $E_{\eta,asym}$
  (red $\triangle$) are displayed in panels (d)-(f) and (j)-(l).
  Electric fields are given in mV/m and all parameters are plotted as
  a function of $L_{Sep}$, the duskward distance in R$_\text{E}$ along
  the separator from the subsolar magnetopause.}
\label{fig::E_Scaling_Base_Simulations}
\end{figure*}

Figure~\ref{fig::E_Scaling_Base_Simulations} displays the measured
reconnection rate $E_{z'}$ (black squares) in mV/m along the separator
in distinct simulations with $\theta_\text{IMF}$ of (a) $180^\circ$,
(b) 150$^\circ$, (c) 120$^\circ$, (g) 90$^\circ$, (h) 60$^\circ$, and
(i) 30$^\circ$.  Also displayed are the theoretical asymmetric
reconnection rates $E_{asym}$ (blue diamonds) and $E_{\eta,asym}$ (red
triangles) given by Eqs.~\eqref{eqn::Asymmetric_Reconnection_Rate}
and~\eqref{eqn::Asymmetric_SP_Reconnection_Rate}, respectively.  The
reconnection rates are plotted as a function of $L_{Sep}$.  Note, the
vertical scale is different for different $\theta_\text{IMF}$;
reconnection is faster for southward IMF than northward IMF, as is
well-known.

The comparison between theoretical and measured values for
$\theta_\text{IMF} = 180^\circ$ in panel (a) reveals that the
prediction is exceedingly good; they are almost indistinguishable.
This is consistent with previous results of~\citet{borovsky2008a}
and~\citet{ouellette2014}.  The other clock angle simulations in
panels (b), (c), and (g)-(i) reveal good agreement in the scaling
sense, meaning that all parameters differ by some coefficient that is
approximately constant along the parts of the separator where
reconnection is strongest. While the scaling is strong for all
simulations, a comparison of absolute quantities shows that the
quantitative agreement becomes progressively worse as the clock angle
decreases.
    
As a way of quantifying the discrepancy between the measured
reconnection rate and the predictions, the percent error between the
measured reconnection rate $E_{z'}$ and the generalized asymmetric
reconnection rate $E_{asym}$ is calculated as a function of the
distance along the separator as
%\begin{linenomath*}
\begin{equation}
  \%\ Error = \frac{E_{asym}-E_{z'}}{E_{z'}}\times100,
\label{eqn::Generalized_Percent_Error}
\end{equation}
%\end{linenomath*}
and for the asymmetric Sweet-Parker reconnection rate $E_{\eta,asym}$
as
%\begin{linenomath*}
\begin{equation}
  \left(\%\ Error\right)_\eta =
  \frac{E_{\eta,asym}-E_{z'}}{E_{z'}}\times100.
\label{eqn::Sweet-Parker_Percent_Error}
\end{equation}
%\end{linenomath*}
The percent errors from Eq.~\eqref{eqn::Generalized_Percent_Error} for
simulations with $\theta_\text{IMF}$ of (d) $180^\circ$, (e)
150$^\circ$, (f) 120$^\circ$, (j) 90$^\circ$, (k) 60$^\circ$, and (l)
30$^\circ$ are displayed as blue diamonds; those from
Eq.~\eqref{eqn::Sweet-Parker_Percent_Error} are red triangles. We note
that through the subsolar magnetopause region
($|L_{Sep}|\le5$~R$_\text{E}$), $E$ is significantly different from
zero and the percent error is relatively constant with distance along
the X-line.  This implies the agreement is good in the scaling sense.

However, there is a trend that the percent error gets larger for
smaller $\theta_\text{IMF}$ for both comparisons.  The percent errors
at the subsolar point for all simulations are given in
Table~\ref{table::Subsolar_Percent_Differences}. The dependence is
described fairly well as the percent error being proportional to
$-\cos\theta_\text{IMF}$ (not shown).  This suggests that there is a
systematic effect causing an offset that increases with decreasing
$\theta_\text{IMF}$.  One possible explanation is it could simply be a
systematic effect in our algorithm to find $\delta$ or $L$ or that the
plane of reconnection is not normal to the X-line for oblique IMF.  It
could also be physical, such as being related to the underlying
assumption of the applicability of the 2D asymmetric reconnection
theory to the magnetopause.

\begin{table*}[b]
  \caption{Percent differences between the measured and predicted
    values of the reconnection rate $E_{z'}$ at the subsolar point.}
  \label{table::Subsolar_Percent_Differences}
  \begin{tabular}{c | c c c c c c | c}
    \hline & 180$^\circ$ & 150$^\circ$ & 120$^\circ$ & 90$^\circ$ &
    60$^\circ$ & 30$^\circ$ & 120$^\circ$, $\psi=15^\circ$\\\hline $\%
    Error$ & $2.67$ & $-3.54$ & $-15.79$ & $-32.99$ & $-57.62$ &
    $-72.15$ & $-20.80$\\ $\left(\% Error\right)_\eta$ & $5.92$ &
    $3.44$ & $-3.69$ & $-13.79$ & $-31.56$ & $-43.00$ &
    $-6.45$\\\hline Prediction Error & $-3.07$ & $-6.75$ & $-12.56$ &
    $-22.26$ & $-38.07$ & $-51.14$ & $-15.34$\\\hline
  \end{tabular}
\end{table*}
We note that the two curves for $E_{\eta,asym}$ and $E_{asym}$ should
ideally lie on top of each other. However, in these cases there is
some offset between the two. The error between the two predictions at
the subsolar point is calculated using a form similar to
Eq.~\eqref{eqn::Generalized_Percent_Error} and substituting
$E_{\eta,asym}$ for $E_{z'}$, and is given in the last row of
Table~\ref{table::Subsolar_Percent_Differences}. The results
underscore that the differences may be attributed to the algorithm
used to measure plasma parameters.

\begin{figure}[t]
  \centering
  \noindent\includegraphics[width=20pc]{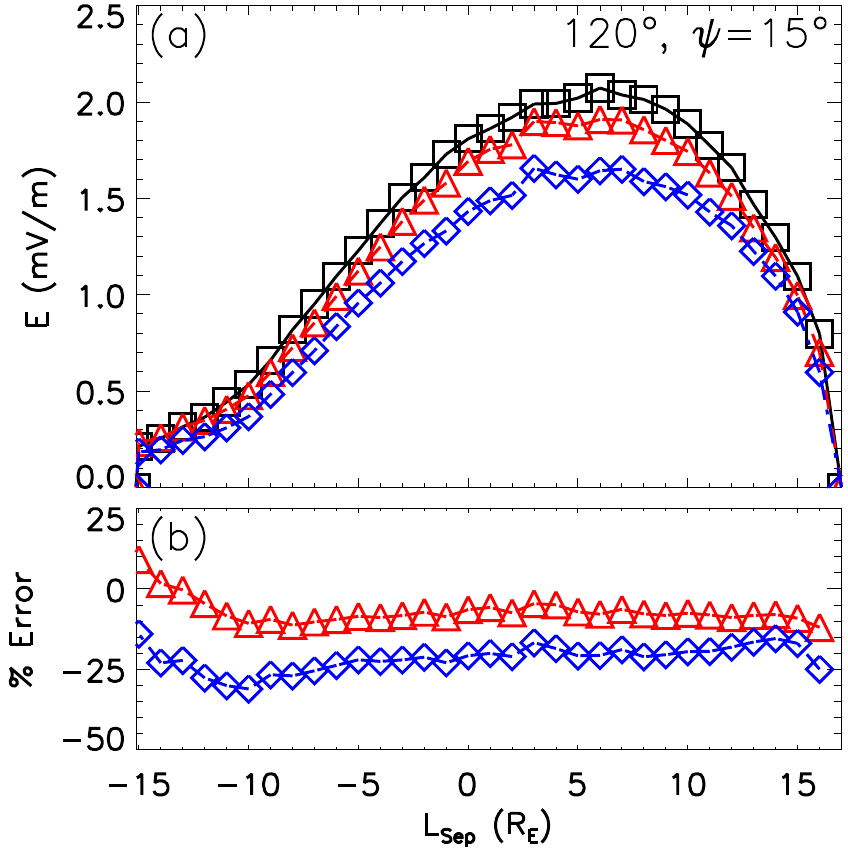}
  \caption{Comparison between the measured reconnection rate at the
    separator with the theoretical asymmetric reconnection rates in a
    simulation with $\theta_\text{IMF}=120^\circ$ and positive dipole
    tilt of $\psi=15^\circ$.  See caption of
    Fig.~\ref{fig::E_Scaling_Base_Simulations} for definitions.}
  \label{fig::E_Scaling_Dipole_Tilt}
\end{figure}

The simulations employed so far all have significant symmetry, so it
is important to do similar comparisons for systems without symmetry.
We therefore include a positive dipole tilt $\psi=15^\circ$ (northern
geomagnetic pole oriented sunward) to break this symmetry.  We use
$\theta_\text{IMF}=120^\circ$ and all solar wind parameters the same
as the previous simulations as a test case.
Figure~\ref{fig::E_Scaling_Dipole_Tilt} displays the reconnection rate
comparison as before for the dipole tilt simulation.  We again see
very good agreement in the scaling sense for both theoretical
reconnection rates.  The percent differences are calculated at the
subsolar magnetopause as before and we find the errors in $E_{asym}$
and $E_{\eta,asym}$ to be $-21\%$ and $-6.5\%$, respectively; these
percent differences are comparable to those of the simulation with the
same IMF clock angle but without any dipole tilt (see
Table~\ref{table::Subsolar_Percent_Differences}).  This suggests that
the prediction is equally successful with a dipole tilt.

Finally, we discuss an important aspect of an additional parametric
test of the theory that could be the cause of confusion in future
studies.  We test smaller IMF strengths of 5 and 2~nT (from 20~nT).
Each simulation uses $\theta_\text{IMF}=120^\circ$ without a dipole
tilt and keeping all other solar wind parameters unchanged.  From
looking at the raw data, it appears that the agreement for the
prediction compared to the measurement is much worse.  The 5 nT
simulation shows limited scaling agreement, and the 2 nT simulation
does not reveal agreement even in the scaling sense.

It is important to realize, however, that the disagreement in this
case is likely numerical.  For smaller IMF strength, the magnetosheath
reconnecting field strength decreases, leading to a larger asymmetry
in magnetic field across the reconnection site.  The larger the
asymmetry, the more the X-line and stagnation point are separated
\citep{cassak2007c}.  As discussed in \citet{cassak2008b}, when the
X-line or stagnation point becomes less than a grid cell from the edge
of the diffusion region, numerical problems arise.  In these two
simulations, the ratios of the magnetospheric reconnecting magnetic
field component to the magnetosheath's are 0.2 and 0.1 for the
$B_\text{IMF}=5$ and 2~nT simulations, respectively.  The X-line is
located much closer to the edge of the magnetosheath current layer,
and the distance between the two falls below our minimum simulation
resolution of 0.125~R$_\text{E}$.  Thus, the reconnection dynamics in
the dissipation region are not sufficiently resolved.  One would need
maximum resolutions of 1/16 and 1/32~R$_\text{E}$ in our
$B_\text{IMF}=5$ and 2~nT simulations, respectively, to properly
resolve the reconnection region substructure.  Care should be taken on
this issue in future studies.

\section{Summary}
\label{section::Summary}

In this study, we have investigated the local properties of magnetic
reconnection at the dayside magnetopause in global MHD
simulations. Previous studies have tested local reconnection theory in
observations and simulations of reconnection at the dayside
magnetopause for predominantly southward IMF, while the present work
systematically finds the 3D extent where reconnection is possible and
tests the 2D theory with oblique IMF and dipole tilt conditions.

The analysis presented here suggests that, up to a scaling factor, the
2D asymmetric reconnection theory accurately predicts the {\it local}
reconnection rate at the dayside magnetopause as a function of the
upstream parameters local to the magnetic X-line for due southward IMF
and without a dipole tilt.  In simulations with oblique IMF, the
analysis techniques are consistent with the scaling of the
reconnection rate from 2D asymmetric reconnection theory, up to an
unspecified constant.  The theory had been confirmed in previous
global MHD simulations at the subsolar magnetopause for due southward
IMF~\citep{borovsky2008a, ouellette2014}.  However, to the best of our
knowledge, the present study is the first of its kind to analyze
reconnection local to the X-line for oblique IMF orientations and
including a dipole tilt.
% The present results suggest that the approach described here by
% measuring the upstream plasma parameters to calculate the local
% reconnection rate along the line of reconnection is appropriate for
% analyzing reconnection at the dayside magnetopause under oblique IMF
% with or without a dipole tilt.

Interestingly, we have found an undetermined constant of
proportionality between the scaling prediction from
\citet{cassak2007c} and the reconnection rates measured for oblique
IMF, with the offset becoming more pronounced for smaller IMF clock
angles.  The cause of the offset is not understood, but it could
either be a systematic effect based on how we determine the upstream
parameters or a systematic limitation of applying the 2D theory to the
magnetosphere.  Future work should be done to make this determination.

This study can be useful to help bridge the gap between reconnection
physics local to the X-line and how the magnetosphere globally reacts
to given input from the solar wind.  This is a core issue in the
recent questions about whether local or global physics control dayside
reconnection.  Specifically, these techniques could be used to
understand how plasmaspheric drainage plumes affect the local and
global reconnection rates at the dayside magnetopause for arbitrary
IMF and magnetospheric dipole tilt.  However, this is beyond the scope
of the present study, and will be important for future work.  It is
hoped that similar tests can be performed with NASA's recently
launched Magnetospheric Multiscale (MMS) mission \citep{burch2015},
which carries instruments with sufficient temporal and spatial
resolution to observe reconnection and has spent a significant amount
of time at the dayside magnetopause.

The approach used here to measure upstream plasma parameters locally
at the X-line should be useful in related work.  There has been an
increase in use of the Hall term in global magnetospheric simulations.
The Hall effect was recently shown to significantly alter the global
dynamics at Jupiter's moon Ganymede, with effects not seen in
resistive MHD~\citep{dorelli2015}.  The inclusion of the Hall term has
profound implications on the local rate of reconnection, with Hall
reconnection being much faster than collisional
reconnection~\citep{birn2001}.  Separators in the Ganymede Hall-MHD
simulations were identified, but they were not used to calculate the
local reconnection rate.  This is because the parallel reconnection
electric field from the Hall term vanishes since
$\mathbf{E}\propto\mathbf{J\times B}$.  The fact that there is good
agreement between the reconnection rate calculated from the parallel
electric field and the prediction based on upstream plasma parameters
in the present study suggest that one can estimate the reconnection
rate in Hall-MHD simulations by measuring the upstream plasma
parameters and calculating the generalized reconnection rate
$E_{asym}$.

The present study employed a few underlying assumptions.  For the
global magnetospheric simulations, we employ a uniform, explicit
resistivity even though Earth's magnetopause is essentially
collisionless; this choice ensures our simulations are well resolved
while reducing the likelihood of flux transfer events
(FTEs)~\citep{russell1978}.  However, recent advances have been made
to trace magnetic separators in simulations with
FTEs~\citep{glocer2016}, so this restriction is not required.

The present study does not take into account the effect of flow shear
at the magnetopause due to the solar wind in
Eq.~\eqref{eqn::Asymmetric_Reconnection_Rate} even though the theory
of asymmetric reconnection with flow shear was recently worked out
\citep{doss2015}.  However, the result of that study is that the flow
shear becomes less important as the magnetosphere/magnetosheath
asymmetries become more significant, so it is possible that the effect
of the flow shear is not very large.  Future work on this is required.

This study also does not take into account that the reconnection
parameters are asymmetric in the outflow direction.  This is
especially true for essentially locations along the X-line where
symmetry is broken: locations away from the subsolar point for all but
due-southward IMF and no dipole tilt, and everywhere when a dipole
tilt is present.  There are very few studies of this effect
\citep{murphy2010}.  This effect undoubtedly is important and should
be taken into account.

Further, the research detailed here uses idealized solar wind
conditions with a few limitations not wholly representative of solar
wind observations.  The present work does not use an IMF $B_x$
component, although it is expected that it affects reconnection in a
similar way to systems with a dipole tilt.  Previous studies found
that under southward IMF orientations, the reconnection site moves
northward for $B_x>0$ and southward when $B_x<0$~\citep{peng2010,
  hoilijoki2014}.  Additionally, our analysis is performed after the
simulations have achieved a quasi-steady-state, which does not capture
the magnetosphere's response to dynamic solar wind
conditions~\citep{laitinen2006, laitinen2007}.  Understanding the
response of Earth's magnetosphere for a broader range of solar wind
conditions is of the utmost importance for realistic space weather
forecasting, and will be the subject of future work.

Finally, this analysis is limited to the dayside portion of the
X-lines.  The X-line extends to the magnetotail where it forms a
closed loop~\citep{laitinen2006, laitinen2007}.  The methodology here
should work for locations extending further poleward of the magnetic
nulls, like those found for northward IMF conditions described
in~\citet{komar2015}, but further research is necessary.

\begin{acknowledgments}
  Support from NSF grant AGS-0953463 and NASA grant NNX10AN08A (PAC),
  and NASA West Virginia Space Grant Consortium (CMK) are gratefully
  acknowledged.  Simulations were performed at the Community
  Coordinated Modeling Center at Goddard Space Flight Center through
  their public Runs on Request system (http://ccmc.gsfc.nasa.gov). The
  CCMC is a multi-agency partnership between NASA, AFMC, AFOSR, AFRL,
  AFWA, NOAA, NSF and ONR. The BATS-R-US Model was developed by the
  Center for Space Environment Modeling at the University of
  Michigan. The analysis presented here was made possible via the
  Kameleon software package provided by the Community Coordinated
  Modeling Center at NASA Goddard Space Flight Center
  (http://ccmc.gsfc.nasa.gov). Software Developers are: M.~M.~Maddox,
  D.~H.~Berrios, and L.~Rastaetter.  The data used to produce the
  results of this paper are publicly available for free from CCMC.
  The authors would like to thank J.~C.~Dorelli, B.~Lavraud,
  Y.~H.~Liu, D.~G.~Sibeck, and F.~D.~Wilder for their insight and
  interesting discussions, and CMK thanks A.~Glocer for his
  mentorship.
\end{acknowledgments}

% \bibliographystyle{agufull08}
% \bibliography{/Users/ckomar/Research/Written_Papers/The_one_bib_to_rule_them_all}
%  \begin{comment}

%\end{comment}

\end{article}
\end{document}